\documentclass[twocolumn,amsmath,amssymb,a4paper,superscriptaddress]{revtex4}

\usepackage{graphicx}

\usepackage{textcomp} 

\begin{document}

\title{Enhancing the optical excitation efficiency of a single self-assembled quantum dot with a plasmonic nanoantenna}

\author{Markus Pfeiffer}

\affiliation{4$^\mathit{th}$ Physics Institute and Research Center SCOPE, University of
Stuttgart, Pfaffenwaldring 57, D-70550 Stuttgart, Germany}

\affiliation
{Max Planck Institute for Solid State Research, Heisenbergstrasse 1, D-70569 Stuttgart,
Germany}

\author{Klas Lindfors}

\affiliation{4$^\mathit{th}$ Physics Institute and Research Center SCOPE, University of
Stuttgart, Pfaffenwaldring 57, D-70550 Stuttgart, Germany}

\affiliation
{Max Planck Institute for Solid State Research, Heisenbergstrasse 1, D-70569 Stuttgart,
Germany}

\author{Christian Wolpert}

\affiliation{4$^\mathit{th}$ Physics Institute and Research Center SCOPE, University of
Stuttgart, Pfaffenwaldring 57, D-70550 Stuttgart, Germany}

\affiliation
{Max Planck Institute for Solid State Research, Heisenbergstrasse 1, D-70569 Stuttgart,
Germany}

\author{Paola Atkinson}

\author{Mohamed Benyoucef}

\author{Armando Rastelli}
\email{a.rastelli@ifw-dresden.de}

\author{Oliver G. Schmidt}

\affiliation{IFW Dresden, Helmholtzstrasse 20,
D-01069 Dresden, Germany}

\author{Harald Giessen}

\affiliation{4$^\mathit{th}$ Physics Institute and Research Center SCOPE, University of
Stuttgart, Pfaffenwaldring 57, D-70550 Stuttgart, Germany}

\author{Markus Lippitz}
\email{m.lippitz@physik.uni-stuttgart.de}

\affiliation{4$^\mathit{th}$ Physics Institute and Research Center SCOPE, University of
Stuttgart, Pfaffenwaldring 57, D-70550 Stuttgart, Germany}

\affiliation
{Max Planck Institute for Solid State Research, Heisenbergstrasse 1, D-70569 Stuttgart,
Germany}

\date{\today}


\begin{abstract}
We demonstrate how the controlled positioning of a plasmonic nanoparticle
modifies the photoluminescence of a single epitaxial GaAs quantum dot. The
antenna particle leads to an increase of the luminescence intensity by about a
factor of eight. Spectrally and temporally resolved photoluminescence
measurements prove an increase of the quantum dot's excitation rate. The
combination of stable epitaxial quantum emitters and plasmonic nanostructures
promises to  be highly beneficial for nanoscience and quantum optics.

\end{abstract}

\maketitle

Coupling optical quantum emitters to plasmonic nanostructures is a particularly topical
area of quantum optics of single nanoobjects as well as plasmonics~\cite{chang:2006,akimov:2007,fedutik:2007}. The localized nature of the electromagnetic field associated
with particle plasmons is equivalent to a small effective mode volume of the
optical field, leading to an increase in the light-matter coupling constant. Many fascinating experiments coupling emitters and plasmons have been proposed or performed, such as enhanced spontanous emission into free space~\cite{shimizu:2002,liu:2003,farahani:2005,mertens:2006,tam:2007,ringler:2008,bek:2008,kinkhabwala:2009,schietinger:2009}, spatially directed emission by an optical nanoantenna~\cite{taminiau:2008}, or the use of the emitter's nonlinearity as an optical transistor~\cite{chang:2007}. All these experiments require a quantum emitter that is stable in position and emission rate, and well defined in emission frequency and orientation of its dipole axis relative to the plasmonic structure.

Epitaxially grown self-assembled semiconductor quantum dots have proven to be bright and non-blinking single photon sources~\cite{michler:2000}, fulfilling all the above requirements. However, because of the high refractive index of the semiconductor substrate ($n \approx 3.5$ for GaAs), the coupling of the quantum dot to the free-space optical field is weak.  A plasmonic nanoantenna close to a quantum dot promises relief of this restriction.

In this Letter we demonstrate to our knowledge for the first time the controlled coupling of a \textit{single} epitaxial quantum dot to a plasmonic nanoantenna. Excellent control over the coupling is a prerequisite for realizing more complex plasmonic devices making use of several single emitters, where one cannot rely anymore on  random placement of emitters relative to the plasmonic structures. To achieve this degree of control,  emitters can be chemically conjugated to the plasmon resonant structure~\cite{tam:2007,ringler:2008}. Alternatively atomic force microscopy (AFM) based
nanomanipulation can be used to approach emitter and metal
nanostructure~\cite{bek:2008,schietinger:2009}. However, especially the orientation of the emitter's dipole axis remains difficult to control. The epitaxial quantum dots used in this work have two energetically almost degenerate transitions with well defined,
orthogonal dipole moments which are parallel to the sample surface and are oriented along specific crystal directions~\cite{plumhof:10}. Placing of the plasmonic structure in a well-defined distance to the quantum dot is straight forward as the dot's position is visible by AFM and electron microscopy due to a characteristic structure on the sample surface. Our exclusively solid-state approach is very  well
suited for developing nanometer-scale optical circuits for future photonic networks.

The quantum dots in our sample are lens-shaped GaAs inclusions of about 10 nm
diameter and 5 nm height embedded in a AlGaAs barrier of larger band-gap. The two optically active dipole transitions of the neutral excitons show splittings of less than 90 $\mu$eV which can be ascribed to slight dot-shape anisotropies~\cite{plumhof:10}. With a ground state exciton emission
wavelength around 760~nm, these quantum dots are well suited for Si-based single
photon detectors. The quantum dots were grown using molecular beam epitaxy. In a
first step a template of nanoholes was created by arsenic debt epitaxy on a GaAs
(100) substrate~\cite{gonzalez:2009}. During further growth steps the sample topography is preserved and the holes are visible by atomic force microscopy on the sample
surface (\ref{fig:nanoman}a). The quantum dots are formed by first
depositing the AlGaAs barrier material followed by a layer of GaAs. During this
growth step, GaAs accumulates in the holes and forms the quantum dots, while
outside the holes only a thin GaAs wetting layer remains~\cite{plumhof:10,wang:2009}. A
detailed description of sample growth is given in the supplementary material.
The quantum dots are located 10~nm beneath the sample surface. The shallow depth
of the emitters enables near-field coupling with plasmonic structures placed on
the sample surface. Here we use gold nanoparticles positioned above the quantum
dots to act as optical antennas. A schematic of the studied structure is shown
in  \ref{fig:nanoman}b.

\begin{figure}
	\centering
\includegraphics{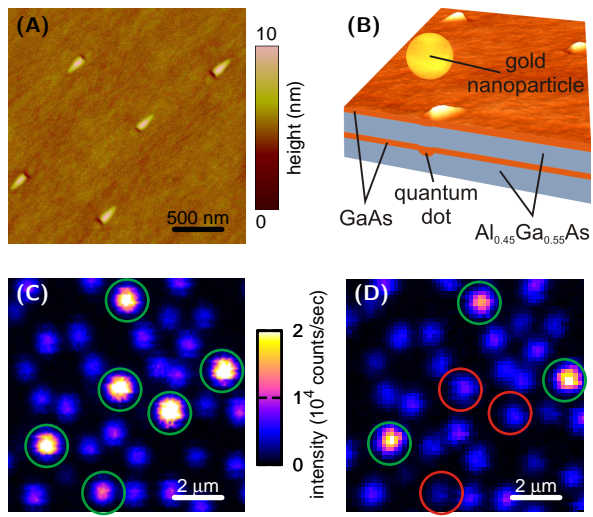}
 \caption{ (a) Single quantum dots appear as a characteristic dip-peak structure
in the topography of the sample surface. (b) Schematic of the studied structure
(not to scale). Here the gold nanoparticle (diameter 90 nm) has not yet been
positioned on top of the quantum dot, which is 10 nm below the sample surface.
(c) Spectrally integrated photoluminescence image of the sample. Each bright spot corresponds to the
emission of a single quantum dot. Emitters decorated with a gold nanoparticle
are marked by green circles. (d) Photoluminescence image of the same sample
region as in (c) after removing the gold particles from the three emitters marked
by red circles.} \label{fig:nanoman}
\end{figure}

The quantum dots can be excited resonantly, i.e., by absorption of a photon by
the quantum dot, or non-resonantly by carriers that are created outside the dot
and diffuse into it. Here we excite the quantum dot non-resonantly by creating
electrons and holes in the barrier and wetting layer using an optical parametric
oscillator. The oscillator produces 400~fs pulses with a photon energy that can
be tuned between 1.8 and 2.4~eV. The photoluminescence is collected  in a home-built
low-temperature confocal laser scanning microscope. The spatial resolution is
limited by the objective's numerical aperture (NA = 0.7) to approximately
600~nm. The photoluminescence can be spectrally analyzed using a spectrometer
equipped with a charge coupled device camera. The spectral resolution of the
setup is around 100 $\mu$eV. For time-resolved measurements we employ
time-correlated single photon counting with a temporal resolution of
approximately 30~ps, which is much shorter than the measured exciton decay time of about
350 ps. All the optical measurements were performed at a temperature of 10~K.

As optical antennas we use spherical gold nanoparticles with a diameter of 90~nm
as specified by the manufacturer (Nanopartz Inc.). The nanoparticles were
deposited on the sample surface by drop casting from an aqueous solution. The
sample had been patterned with markers so that the same area can be studied with AFM and optical microscopy. An AFM was used to identify regions
of the sample with a low concentration of gold particles. Ten of the nanoparticles were moved on top of nearby quantum
dots using the AFM tip. The optical properties of nanoantennas were
characterized by recording the scattering spectrum of single antennas using
dark-field microspectroscopy~\cite{schultz:2000,sonnichsen:2002}. Details of the
dark-field spectroscopy are given in the supplementary material.

\ref{fig:nanoman}c shows a raster-scanned photoluminescence image of the region of the
sample containing antenna-coupled emitters. For each bright, diffraction-limited
spot in the luminescence image one can find a corresponding surface depression
in an AFM image of the same area of the sample. The majority of the quantum dots
are comparable in brightness. The emitters decorated with a gold nanoparticle,
however, appear markedly brighter as seen in \ref{fig:nanoman}c. To exclude
a coincidence, we removed the particle from three quantum dots (marked by red
circles in \ref{fig:nanoman}d) after characterizing their optical
properties. The photoluminescence of these three quantum dots changed to
approximately the ensemble average value.

\begin{figure}
	\centering
\includegraphics{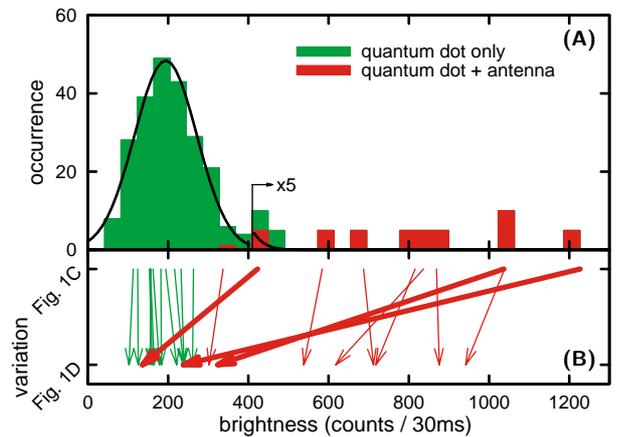}
\caption{(a) Brightness distribution for 228 reference quantum dots (green) and
the ten antenna-enhanced emitters (red). The solid line is a Gaussian fit to the
data of the reference quantum dots. (b) Change in the brightness of emitters
from which the antenna was removed (thick red arrows). The brightness of the
reference quantum dots (a few shown as thin green arrows) as well as the unmodified quantum
dot-nanoparticle complexes (thin red arrows) remain at their previous level.}
\label{fig:histo}
\end{figure}

To quantify the influence of the nanoparticle on the emission properties of the
quantum dots, we analyzed the photoluminescence of 228 emitters by fitting a
two-dimensional Gaussian function to each emission spot in the luminescence
image. The thus obtained brightness distribution is shown in
\ref{fig:histo}. The center of the distribution was determined as 194
counts/30 ms (6470 counts/s) by fitting a Gaussian function to the histogram. The
brightness of the 10 emitters coupled to a nanoantenna was determined in a
similar fashion. We observe an increase in the photoluminescence intensity of up to a
factor of 6 compared to the ensemble average. All our antenna-coupled quantum
dots have a brightness that is greater than the ensemble average value. One
advantage of our approach is the possibility to compare the properties of
an emitter-antenna complex directly with those of the same single emitter
without the antenna. Here we did this by removing the gold nanoparticle from the
quantum dot for the three dots encircled in red in \ref{fig:nanoman}d. The
change in the photoluminescence intensity is shown in \ref{fig:histo}b. Both the
reference quantum dots and the unmodified emitter-antenna complexes remain at a
similar brightness, as expected. The brightness of the three emitters from which
the antennas were removed decreased to a value well within the ensemble
distribution. For these quantum dots we deduce photoluminescence enhancement
factors of 3.5, 3.7, and 6.1 when compared to their emission without antenna. The spread in the enhancement can be explained by a variation of the relative position of the antennas with respect to the quantum dots.

\begin{figure}
\centering
\includegraphics{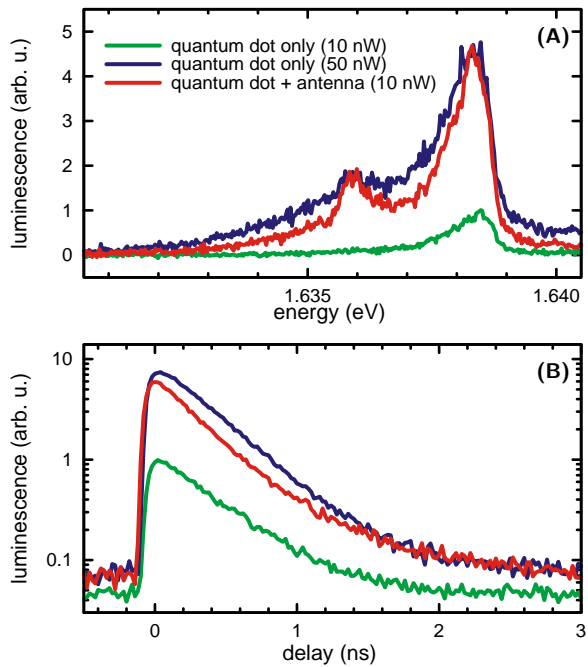}
\caption{(a) Photoluminescence emission spectra of a single quantum dot excited with
photons of energy 2.06 eV: emitter with antenna at 10~nW excitation power (red)
and without antenna at 10~nW (green) and 50~nW (blue) excitation power. (b)
Normalized photoluminescence decay traces for the same quantum dot corresponding to the spectra in a). }
\label{fig:emissiondecay}
\end{figure}

To determine the origin of the enhancement in photoluminescence intensity  let us compare
the emission spectra of the quantum dots with and without antenna as shown in
\ref{fig:emissiondecay}a. We remark that the photoluminescence spectrum
displayed in \ref{fig:emissiondecay}a for the quantum dot coupled to the
antenna was shifted in energy by 1.4~meV to compensate for spectral shifts most likely due
to surface charges in the vicinity of the quantum dot.
At 10~nW excitation power the emission spectrum of the quantum dot without
antenna is dominated by the exciton transition (\ref{fig:emissiondecay}a).
At 50~nW excitation power the exciton line starts to saturate and a second
emission line appears. Compatible with previous reports on GaAs/AlGaAs dots~\cite{wang:2009}, we identify this as one of the trions, i.e., charged exciton, transition from its super-linear excitation power dependence and its lifetime.
The antenna-enhanced quantum dots show already at 10~nW excitation power an
emission spectrum that resembles that of the unenhanced emitter at higher
excitation power (\ref{fig:emissiondecay}a). From this we conclude that
the emitter in the quantum dot-nanoparticle complex experiences an about 5 times higher excitation power density due to the gold particle. The field enhancement around
the plasmonic nanoparticle locally increases the probability that an electron-hole pair
is created. The photoluminescence enhancement as determined from luminescence
images is 6.1 for this emitter. This is in good agreement with the ratio of the
excitation powers for which the emission spectra of the quantum dot with and
without antenna is similar. Also for the other emitters the agreement between
the enhancement values determined with the two methods was good. We can thus
attribute the brightness enhancement predominantly to an increase of the
excitation rate. This is also supported by measurements of spectrally integrated luminescence decay.
In \ref{fig:emissiondecay}b we show the luminescence decay traces
corresponding to the spectra displayed in \ref{fig:emissiondecay}a for 10
nW excitation power and for 50nW with and without antenna, respectively. We observe no significant changes in the decay dynamics due to the antenna.

We note that the emission lines of the antenna-quantum dot complexes appear slightly narrower than the emission of the same dots without antenna at comparable emission intensity. In the usual far-field excitation scheme carriers are generated in a larger volume than in the presence of the antenna. This leads to fluctuating charges, broadening the dot's emission by spectral diffusion.

To further elucidate the effect of the plasmonic antenna on the optical
properties of the quantum dots in our structure we have studied the influence of
the excitation wavelength on the photoluminescence. For non-resonant excitation,
the luminescence rate of the emitters depends on the carrier density in their
vicinity and their diffusion into the quantum dot. Assuming a dipole transition, for low excitation power the
carrier density is proportional to the local intensity of the electromagnetic field. The excitation rate of the quantum dots will thus be maximized when the
excitation photon energy is tuned to the antenna resonance.

\begin{figure}
\centering
\includegraphics{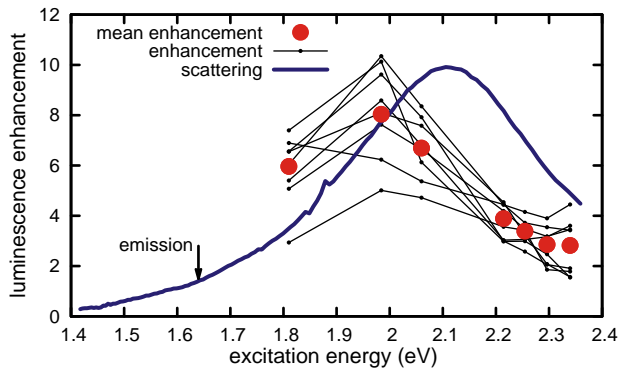}
\caption{Spectral dependence of the luminescence enhancement averaged over 8 
quantum dots decorated with single gold spheres (red filled circles) compared
to the measured dark-field scattering spectrum of a single antenna particle (blue
line). The excitation enhancement spectrum of the individual dots (thin black lines) is very similar after normalizing them to have the same mean enhancement. The latter varies  significantly  from dot to dot (see, e.g., \ref{fig:histo} for the spread at 2.06 eV).
} \label{fig:spectral_enh}
\end{figure}

\ref{fig:spectral_enh} shows the photoluminescence enhancement of the
quantum dots coupled to a nanoparticle antenna as a function of excitation energy for an average excitation power of about 50 nW. The enhancement was determined by comparing the brightness of the
antenna-enhanced emitters to an ensemble of reference quantum dots that were not
coupled to an antenna. We observe a peak in the measured photoluminescence
enhancement at approximately 2.0~eV energy. At the plasmon
resonance the electromagnetic field is enhanced close to the particle and the
excitation rate of electron-hole pairs in the vicinity of the quantum dot is
thus increased. 
The enhancement maximum occurs at a slightly lower energy than the plasmon resonance as seen from the scattering
spectrum shown in \ref{fig:spectral_enh}. This is in agreement with the
report of Bryant~\emph{et al.}~\cite{Bryant:2008} that the near-field resonance occurs red-shifted of the far-field scattering resonance. The emission of the quantum dots is not modified (\ref{fig:emissiondecay}b) as it occurs far off the antenna's plasmon resonance.

In summary, we have introduced single epitaxial GaAs~quantum dots as emitters for plasmonics by demonstrating the controlled enhancement of the excitation efficiency. Placing a gold nanoparticle in the vicinity of the emitter resulted in an increase in the
photoluminescence intensity of about a factor of eight.
In the near-field of the antenna close to the quantum dot more carriers are generated so that  the excitation rate is increased.
Our conclusion is supported by the observation that
the decay dynamics of the emitters is unaltered by the coupling to the antenna.
Varying the wavelength of the incident light clearly demonstrates that the
photoluminescence enhancement originates from the plasmon resonance of the
antenna.

Our scheme can be applied to more complex antennas and more than one emitter.
For example, rod-like plasmon resonant metal particles can be positioned in the
same way as here using AFM. Moreover, the topography of the quantum dots can be
used to first determine their position and subsequently fabricate antenna
structures by electron-beam lithography on top of them. The GaAs quantum dots
provide a stable, solid-state source of photons that, when coupled to plasmonic
nanostructures, forms an attractive system in which to study light-matter
interactions and to use as a basis for nanophotonic applications.

We thank A. Vlandas for assistance with atomic force microscopy, L. Wang, F. Ding and B. Eichler for the growth and characterization of test samples, and the Deutsche Forschungsgemeinschaft DFG for funding (research unit FOR 730).

\maketitle

\section{Supplementary Material}

\subsection{Growth of the quantum dots}

The GaAs quantum dots are grown by molecular beam epitaxy. A low density
($\approx 5\times10^7$~cm$^{-2}$) of self-assembled nanoholes approximately
10~nm deep and 100~nm wide were created by arsenic debt epitaxy in a GaAs (100)
substrate~[S1]. Deposition of 21 repeats of 0.4 monolayers Ga / 0.15 monolayers
GaAs lead to droplets of excess gallium on the surface. A 5 minute interrupt
under excess arsenic at a substrate temperature of 520\,$^\circ$C results in the
formation of nanoholes surrounded by a raised GaAs ring due to the transfer of
arsenic from the substrate to the gallium droplets~[S2]. A 10~nm bottom
Al$_{0.45}$Ga$_{0.55}$As barrier was then grown over the nanoholes, followed by
a 2 nm GaAs layer, an 8 nm top Al$_{0.45}$Ga$_{0.55}$As barrier and a final 2 nm
GaAs capping layer. The substrate temperature was then reduced from
520\,$^\circ$C to approximately 200\,$^\circ$C, and one monolayer of gallium was
deposited to ensure a gallium terminated surface followed by $2\times10^{11}$~cm$^{-2}$ silicon to passivate the surface states~[S3]. During the
growth of the AlGaAs layers the shape of the nanoholes is well preserved.
However, due to the greater migration length of gallium compared to aluminium,
there is accumulation of gallium at the bottom of the nanohole during deposition
of the GaAs layer leading to GaAs quantum dot formation~[S4]. At the end of the
growth, there is still a 2 nm deep depression in the sample surface marking the
position of the quantum dot in the self-assembled nanohole. The elevated feature
corresponds to the GaAs ring while the depression is at the position of the
quantum dot.

\subsection{Dark-field microspectroscopy}

Dark-field scattering spectra of the gold nanoparticles were obtained using a
home-built dark-field microspectroscopy setup. Light from a halogen lamp was
used to illuminate the sample through a dark-field objective (Olympus MPlanFL N
100$\times$/0.90 BD). The sample was positioned using a piezoelectric
positioning system. The light back-scattered by a nanoparticle was collected
with the microscope objective and directed to a 500~mm focal length grating spectrometer
equipped with a liquid nitrogen cooled charge coupled device camera. A reference
spectrum was collected from an area on the sample without nanoparticles and
subtracted from the scattering spectrum of the particle. Finally, to compensate
for the wavelength dependent transmission of the optical train, the emission
spectrum of the lamp, and the wavelength dependent sensitivity of the camera, a
bright-field reflection spectrum of the sample from an area without
nanoparticles was measured. This spectrum was used to normalize the dark-field
spectra to obtain the scattering spectra of the nanoparticles.

\subsection{References}
\setlength{\parindent}{0pt}

[S1] Alonso-Gonz\'{a}lez,~P.; Mart\'{\i}n-S\'{a}nchez,~J.; Gonz\'{a}lez,~Y.; Al\'{e}n,~B.; Fuster,~D.; Gonz\'{a}lez,~L.~\textit{Crystal~Growth~\&~Design}~\textbf{2009},~\textit{9}, 2525.

[S2] Heyn,~Ch.; Stemmann,~A.;~Hansen,~W. \textit{Appl.~Phys.~Lett.}~\textbf{2009},~\textit{95}, 173110.

[S3] Hasegawa,~H.; Akazawa,~M.~\textit{Appl.~Surf.~Sci.}~\textbf{2008},~\textit{255}, 628.

[S4] Rastelli,~A.; Stufler,~S.; Schliwa,~A.; Songmuang,~R.; Manzano,~C.; Costantini,~G.; Kern,~K.; Zrenner,~A.; Bimberg,~D.; Schmidt,~O.~G.~\textit{Phys.~Rev.~Lett.}~\textbf{2004},~\textit{92}, 166104.

\end{document}